\documentclass[showkeys,showpacs,aps,preprint]{revtex4}

\begin{document}

\title{The Cosmological Perturbation of Braneworld with An
Anisotropic Bulk}

\author{Dan-Tao Peng \thanks{dtpeng@ustc.edu.cn or
dtpeng@nwu.edu.cn}\\
The Interdisciplinary Center for Theoretical Study, \\
University of Science \& Technology of China, \\
Hefei, 230026, P. R. China\\
Institute of Modern Physics, Northwest University, \\
Xi'an, 710069, P. R. China\\}

\begin{abstract}
The braneworld cosmological model was constructed by embedding a
3-brane into a higher dimensional bulk background geometry and the
usual matter of our universe was assumed to be confined in the
brane while the gravity can propagate in the bulk. By considering
that the bulk geometry is anisotropic, after reviewed the
separable solution for the bulk metric and the result that the
anisotropic property of the bulk can support the perfect fluid
kind of the matter in the brane, we develop a formalism of the
cosmological perturbation for this kind of anisotropic braneworld
model. As in isotropic case, we can also decompose the
perturbation into scalar, vector and tensor modes, we find that
the formalism for the anisotropic braneworld cosmological
perturbation are very different from the isotropic case. The
anisotropic effect can be reflected in the tensor modes which
dominated the cosmological gravitation waves. Finally, we also
discussed the perturbed Einstein equations governed the dynamics
of the bulk geometry and the brane with the junction conditions.
\end{abstract}



\preprint{USTC-ICTS-05-02}

\maketitle

\section{Introduction}

The braneworld model with large extra dimensions has been studied
for several years (for reviews see \cite{Langlois, BB, Maartens}).
In braneworld cosmological scenarios, the ordinary Standard Model
(SM) of particle physics matters of our "universe" are assumed to
be confined in a lower dimensional brane which can move in a
higher dimensional bulk spacetime, while gravity can propagate in
the bulk. The braneworld model was first proposed by Randall and
Sundrum \cite{RS1, RS2} as a possible solution to the hierarchy
problem between the weak and Planck scales. In view of cosmology,
the simplest braneworld model can be constructed by embedding the
4-D Friedmann-Robertson-Walker (FRW) cosmologies in a 5-D bulk
spacetime. Here the FRW cosmology describes a homogenous and
isotropic brane in which the matter contents are assumed as
perfect fluid and the brane Friedmann equation can be derived out
directly \cite{BDL, BDEL, Kraus, Ida, MSM}. The homogenous and
isotropic symmetries of the brane impose that the static bulk is
necessarily the Schwarzchild-Anti-de Sitter spacetime \cite{BCG}.

The astronomical observations suggest that the homogeneous and
isotropic cosmological model is adequate to describe our universe.
Although our universe seems homogeneous and isotropic today, it
does not mean that the early era of our universe is necessarily
homogeneous and isotropic and there are no observational data
guaranteeing the isotropy in the era prior to the recombination.
It was argued theoretically that the existence of an anisotropic
phase can approach an isotropic one \cite{Misner}. For this
reason, as in the standard 4-D cosmology \cite{Saha}, it is worth
studying the direct generalization of the homogeneous and
isotropic braneworld cosmology, the homogeneous but anisotropic
brane worlds \cite{MSS, CS1, CS2, AL, ACL, Paul, HM, SVF, Frolov,
CMMS, BH, CH, Halpern, Toporensky, BD, FLSZ}.

It is well known that the most information of the early stage of
our universe in present, including the origin of the large scale
structure of our universe is contained in the cosmological
perturbations. The cosmological perturbation provide us with a
window to understand the early universe. So it is very important
to analyze the behave of the cosmological perturbation in the
context of braneworld cosmology. That is to say it would be
crucial to test whether the predictions in the braneworld
cosmological perturbation are compatible with current astronomical
observations, especially the observations of Cosmic Microwave
Background Radiation (CMBR). It would also be very important to
check whether the usual generation mechanism of cosmological
perturbation would be still valid in brane cosmology (see Ref.
\cite{MWBH}).

Recently, in \cite{FLSZ}, Fabbri, Langlois, Steer and Zegers gave
out an explicit vacuum bulk solution for the spatially anisotropic
braneworld cosmology with a negative cosmological constant. Then
by embedding the $Z_{2}$ symmetric branes in this background, they
found that for some bulk solutions, it is possible to embed a
brane with perfect fluid in the bulk though the brane energy
density and pressure are completely determined by the bulk
geometry. That is to say, in the context of brane cosmology, the
observed homogenous and isotropic matter distribution of our
universe does not mean that our universe is necessarily isotropic,
but may be anisotropic. For this reason, since the most
information of the early era of our universe is contained in the
cosmological pertrubation, it is worth for us studying the
anisotropic effect in the cosmological perturbation.

In this paper we will develop a formalism of the linear
cosmological perturbation for the anisotropic braneworld models.
First in section 2 we review the cosmological solution for the
anisotropic braneworld obtained in \cite{FLSZ}. Then in section 3
by using the method proposed in \cite{Langlois2}, we will first
give out the the formalism that can describe the evolution of the
linear perturbation in the bulk for the anisotropic braneworld
cosmology. By taking into acount the dynamics of the brane which
was governed by the Einstein equations with the juntion conditions
between the bulk and the brane matters, the relations between the
perturbation of bulk metric and the perturbation of the matters in
the brane are consisdered in section 4. Finally, in the last
section we give some conculusions and remarks.

\section{The braneworld universe with an anisotropic bulk}

The metric ansatz for the braneworld universe with an anisotropic
bulk (Bianchi I type (BI) braneworld universe) can be constrcuted
by generalizing the usual FRW isotropic braneworld directly to the
spatially anisotropic case \cite{FLSZ}
\begin{equation} \label{metric}
d s^{2} = - n^2(t, y) d t^{2} + \sum_{i = 1}^{3} a_{i}^{2}(t, y)
(d x^{i})^{2} + d y^{2},
\end{equation}
where the coodinates $x^{i}$ dennotes the original three spatial
dimensions and $y$ is the coordinate of extra dimension. Unlike
the FRW braneworld cosmology which has the same scale factor for
all spatial directions, the BI braneworld universe is constructed
by assigning different scale factors for different spatial
directions, thereby introducing the anisotropy to the system. Here
the different scale factors are represented by three different
functions $a_{i}(t, y)$, $(i = 1, 2, 3)$ respectively.

For simplicity, the function form of the scale factors can be
selected as
\begin{equation} \label{aA}
n(t, y) = a_{0}(t, y) = e^{A_{0}(t, y)}, \quad a_{i}(t, y) =
e^{A_{i}(t, y)},
\end{equation}
Then by introducing the the average scale factor $a \equiv
\exp(\sum_{i} A_{i})$ with the defination
\begin{equation} \label{A}
a \equiv e^{A}, \quad A = \frac{1}{3} \sum_{i = 1}^{3} A_{i},
\end{equation}
the anisotropic property of the bulk metric can be represented by
a vector ${\bf N}$ which is defined as
\begin{equation} \label{N}
N_{i} \equiv A_{i} - A
\end{equation}
and satisfy
\begin{equation} \label{N_constraints}
\sum_{i} N_{i} = 0.
\end{equation}

\subsection{Vacuum solutions for the bulk Einstein equations}

By using this separation of the scale factor to the isotropic and
anisotropic parts in the metric (\ref{metric}), the vacuum
Einstein equations in the bulk with a negative cosmological
constant $\Lambda \equiv - 6/\ell^{2}$
\begin{equation}
R_{A B} = \frac{2}{3} \Lambda g_{A B} \quad (A, B = 0, \cdots, 4)
\end{equation}
can then be reexpressed in terms of the isotropic and anisotropic
quantities defined by (\ref{A}) - (\ref{N}) as
\begin{eqnarray} \label{Einstein1}
A_{0}^{\prime \prime} + A_{0}^{\prime} (A_{0}^{\prime} + 3
A^{\prime}) + e^{- 2 A_{0}} \left [ - 3 \ddot{A} - 3 \dot{A}^{2} -
{\bf \dot{N}}^{2} + 3 \dot{A}_{0} \dot{A} \right ] =
\frac{4}{\ell^{2}} \\
\label{Einstein2}
A^{\prime \prime} + A^{\prime} (A_{0}^{\prime} +
3 A^{\prime}) + e^{- 2 A_{0}} \left [ - \ddot{A} - 3 \dot{A}^{2} +
\dot{A}_{0} \dot{A} \right ] = \frac{4}{\ell^{2}} \\
\label{Einstein3}
H^{\prime} e^{- 4 A} + \frac{1}{3} {{\bf
N}^{\prime}}^{2} A^{\prime} + \frac{1}{3} e^{- 2 A_{0}} \left (
{\bf \dot{N}}^{2} A^{\prime} - 2 \dot{A} \dot{\bf N} \cdot {\bf
N}^{\prime} \right ) = \frac{4}{\ell^{2}} \\
\label{Einstein4}
\dot{H} e^{- 4 A} - \frac{1}{3} {{\bf
N}^{\prime}}^{2} \dot{A} - \frac{1}{3} e^{- 2 A_{0}} \dot{\bf
N}^{2} \dot{A} + \frac{2}{3} A^{\prime} \dot{\bf N} \cdot {\bf
N}^{\prime} = \frac{4}{\ell^{2}}
\end{eqnarray}
and
\begin{equation}
\left [ {\bf N}^{\prime} e^{3 A + A_{0}} \right ]^{\prime} = \left
[ \dot{\bf N} e^{3 A - A_{0}} \right ]^{\cdot}
\end{equation}
where a prime denotes the derivatives with respect to the extra
dimensional coordinate $y$, and a dot one with respect to the time
$t$. The quantity $H$ is defined by
\begin{equation}
H \equiv e^{4 A} \left ( {A^{\prime}}^{2} - e^{- 2 A_{0}}
\dot{A}^{2} \right ).
\end{equation}
When ${\bf N} = 0$, we can recover the vacuum Einstein equations
for the isotropic braneworld models \cite{BDEL}.

The difficulties of integrating above equations indicate that in
order to find the explicit solutions we have to consider some
special cases. Here we assume that the metric is seperable:
\begin{equation}
A_{\mu}(t, y) = \alpha_{\mu}(t) + {\cal A}_{\mu}(y),
\end{equation}
then an analytic solution can be obtained by integrating the
Eintein equations (\ref{Einstein1}) - (\ref{Einstein4}) directly.
The solution is \cite{FLSZ}
\begin{equation} \label{metric_solution}
d s^{2} = \sinh^{1/2} \left ( 4 y / \ell \right ) \left [ -
\tanh^{2 q_{0}} \left ( 2 y / \ell \right ) d t^{2} + \sum_{i}
\tanh^{2 q_{i}} \left ( 2 y / \ell \right ) t^{2 p_{i}} \left ( d
x^{i} \right )^{2} \right ] + d y^{2}
\end{equation}
where the seven coeffecients $q_{\mu}$ and $p_{i}$ satisfy the
following constaints
\begin{equation}
\sum_{\mu} q_{\mu} = 0, \quad \sum_{\mu} q_{\mu}^{2} =
\frac{3}{4}, \quad \sum_{i} p_{i} = 1, \quad \sum_{i} p_{i}^{2} =
1, \quad \sum_{i} q_{i}(p_{i} + 1) = 0.
\end{equation}
This metric "mixes" the 5-D static solution in \cite{CMMS} and the
well-known 4-D Kasner solution.

\subsection{The embedded brane and the junction conditions}
\label{brane&juncitoncondition}

When an infinite thin brane was embedded in the bulk, the brane
dynamics was governed by the bulk Einstein equations with the
junction conditions. Here the junction conditions can be obtained
by imposing a $Z_{2}$ symmetry on the brane configurations.

For an embedded brane in the bulk, in case of the $Z_{2}$
symmetries, the Israel junction conditions can be expressed as
\begin{equation} \label{junction}
K^{a}_{b} = - \frac{\kappa^{2}}{2} \left ( T^{a}_{b} - \frac{T}{3}
\delta^{a}_{b} \right ),
\end{equation}
where $T^{a}_{b}$ denote the energy-momentum stress tensor on the
brane and $K^{a}_{b}$ is the extrinsic curvature on one side of
the brane which is defined by
\begin{equation}
K_{a b} = X^{A}_{a} X^{B}_{b} D_{A} n_{B} = \frac{1}{2} \left [
g_{A B} \left ( X^{A}_{a} \partial_{b} n^{B} + X^{A}_{b}
\partial_{a} n^{B} \right ) + X^{A}_{a} X^{B}_{b} n^{C} g_{A B, C}
\right ]
\end{equation}
where $D_{A}$ is the covariant derivatives associated with the
bulk metric $g_{A B}$, $n^{A}$ is the unit vector normal to the
brane. Here the geometry of the brane is defined by its embedding
in the bulk, i.e. $X^{A} = X^{A}(x^{a})$, where $x^{a}$ are the
intrinsic coordinates on the brane and $X^{A}_{a} \equiv
\partial X^{A} / \partial x^{a}$.

For BI braneworld cosmological models, the embedded brane must
respect the BI symmetries. The anisotropic property of the bulk
imply that the brane energy-momentum stress tensor is necessarily
to take the form:
\begin{equation}
T^{a}_{b} = {\rm diag} \left ( - \rho, P_{1}, P_{2}, P_{3} \right
),
\end{equation}
Here in different spatial directions, the pressure of the matter
takes different value and all off-diagonal elements in the stress
tensor are necessarily zero, i.e. the stress tensor of the brane
matter is still anisotropic. When all $P_{i}$ $(i = 1, 2, 3)$ take
the same value, i.e. $P_{1} = P_{2} = P_{3}$, this stress tensor
form go back to the isotropic case, i.e. the perfect fluid form.

As the metric form of the anisotropic bulk, the brane pressure can
also be separated into isotropic and anisotropic part:
\begin{equation}
P_{i} = P + \pi_{i}
\end{equation}
where the anisotropic property of the brane pressure are
represented by the vector $\pi_{i}$ which satisfy: $\sum_{i}
\pi_{i} = 0$.

By this separation, the spatial components of the junction
condition (\ref{junction}) can also be seperated into an isotropic
part:
\begin{equation}
e^{- A_{0}} \dot{y}_{\rm b} \dot{A}|_{\rm b} + \sqrt{1 +
\dot{y}_{\rm b}^{2}} A^{\prime}|_{\rm b} = \frac{\kappa^{2}}{6}
\rho,
\end{equation}
and an anisotropic part
\begin{equation} \label{anisotropic_constraint}
e^{- A_{0}} \dot{y}_{\rm b} \dot{N}_{i}|_{\rm b} + \sqrt{1 +
\dot{y}_{\rm b}^{2}} N^{\prime}_{i}|_{\rm b} =
\frac{\kappa^{2}}{2} \pi_{i}
\end{equation}
This last equation tells us that the anisotropic property of the
bulk and the stress on the brane must be consistent.

It should be noticed here that the brane position in the extra
dimension $y_{\rm b}$ is not assumed to be fixed. In this sense,
the coordinate system is not Gaussian Normal.

For the separable metric solution (\ref{metric_solution}), it is
easy to find that when
\begin{equation}
q_{0} = \pm \frac{\sqrt{3}}{4},
\end{equation}
the brane can support perfect fluid type of matter, i.e. $\pi_{i}
= 0$. That is to say that in an anisotropic background bulk
geometry, we still can embed in a brane with isotropic perfect
fluid as matter. This implies that in the sense of braneworld
cosmology, the isotropic brane matter distribution does not imply
that the background bulk geometry is necessarily isotropic, an
anisotropic background bulk geoemtry is also possilble, while the
trajectory of brane in the bulk can no longer be choosed
arbitrarily in constrast with the isotropic case, it must satisfy
the first integration of (\ref{anisotropic_constraint}). The type
of the perfect fluid matter in the brane must also be compatible
with the given bulk geometry.

\section{The metric perturbation}

In the previous section we have reviewed the results obtained in
\cite{FLSZ}, we see that in an anisotropic background bulk
geometry, it is also possible to embed an brane with perfect fluid
type of matter. That is to say in the context braneworld
cosmology, the observed homogeneous and isotropic matter
distribution can not tell us that the background geometry is
necessarily isotropic.

In this section we will develop a linear cosmological perturbation
formalism for the anisotropic bulk system. As in \cite{Langlois2}
where the formalism for the usual isotropic braneworld
cosmological perturbation was given out, here we still work in the
Gaussian Normal (GN) system of coordinates adapted to the embedded
brane (\ref{metric}) in which the brane is localized at $y = 0$.

The most general perturbed form for the metric (\ref{metric}) can
be write as
\begin{eqnarray} \label{perturbed_metric}
d s^{2} & = &  \left ( g_{A B} + h_{A B} \right ) d x^{A} d x^{B}
\nonumber\\
& = & - n^2 ( 1 + 2 \psi ) d t^2 + 2 \sum_{i} B_{i} d x^{i} d t +
\sum_{i, j} a_{i} a_{j} \left ( \delta_{i j} + \hat{h}_{i j}
\right ) d x^{i} d x^{j} + d y^{2}.
\end{eqnarray}
Due to the complexity of the index system for the anisotropic bulk
metic, in the context following we will always write the summation
symbol explicitly. As in \cite{Bardeen} (see also \cite{MFB}), the
perturbation field in the perturbed metric
(\ref{perturbed_metric}) can also be classified further into
scalar, vector and tensor quantities by according their
transformation properties with respect to the three spatial
directions.  So we have
\begin{equation}
B_{i} = \nabla_{i}B - S_{i}
\end{equation}
and
\begin{equation}
\hat{h}_{i j} = 2 R \delta_{i j} + \nabla_{i} \nabla_{j} E +
\nabla_{i} F_{j} + \nabla_{j} F_{i} + s_{i j}
\end{equation}
where $\nabla_{i}$ denotes the covariant derivatives associated
with the $3 \times 3$ anisotropic spatial metric.  In this
definition of the perturbed metric, $\psi$, $B$, $R$ and $E$ are 4
scalar perturbation fields, $S_{i}$ and $F_{i}$ are 2
divergence-free 3-vectors perturbation fields and $s_{ij}$ is a
transverse and traceless 3-tensor perturbation field. Here we
should notice that different with the usual isotropic case, the
index of the vector and tensor perturbation fields can not upper
and lower directly by the anisotropic spatial metric, but the
index of their product with the anisotropic scale factor together
can be upper and lower by the anisotropic spatial metric. This is
because that for the perturbed anisotropic bulk metric, all the
spatial perturbed fields have to be multiplied by the anisotropic
scale factors. Then by this definition of the perturbed metric,
the divergence-free constraint conditions for the vector
perturbations fields $S_{i}$ becomes
\begin{equation}
\sum_{i} \nabla^{i} S_{i} = \sum_{i} \frac{1}{a_{i}^{2}} S_{i, i}
= 0.
\end{equation}
While the divergence-free constraint conditions for the vector
perturbation fields $F_{i}$ are same as in the isotropic case
\begin{equation}
\sum_{i} a_{i}^{2} \nabla^{i} F_{i} = \sum_{i} F_{i, i} = 0.
\end{equation}
The transverse traceless constraint conditions for the tensor
perturbation fields $s_{i j}$ become
\begin{equation}
\sum_{i} s_{i i} = 0,
\end{equation}
and
\begin{equation}
\sum_{i} a_{i} a_{j} \nabla^{i} s_{i j} = \sum_{i}
\frac{a_{j}}{a_{i}} s_{i j, i} = 0.
\end{equation}
Here a comma denotes the differentiation with respect to the
corresponding spatial coordinates. From these constraints
conditions we know that the vector metric perturbation of the bulk
has 4 degree of freedom, while the tensor perturbation has 2
degree of freedom.

\subsection{Gauge transformation of the metric perturbation}

In different coordinate systems, the metric perturbation defined
above can be different quantitatively, i.e. there exist some gauge
freedoms for the metric perturbation. In order to distinguish the
gauge effect and physical degrees of freedom, we need to study the
coordinate transformation effect on all metric perturbation fields
defined above.

The infinitesimal changes of coordinates
\begin{equation}
x^{A} \rightarrow \tilde{x}^{A} = x^{A} + \xi^{A},
\end{equation}
can induce the transformations for the bulk metric perturbations:
\begin{equation} \label{metricperturbationtransformation}
h_{A B} \rightarrow \tilde{h}_{A B} = h_{A B} - D_{A} \xi_{A} -
D_{B} \xi_{A},
\end{equation}
here $D_{A}$ is the covariant derivatives associated with the
unperturbed bulk metric.

Now we parametrize the infinitesimal coordinate transformation by
the vector $\xi^{A} = (\xi^{0}, \xi^{i}, \xi^{5})$. As discussed
in \cite{Langlois2} for the isotropic case, in a GN coordinate
system, the metric perturbation components related to the extra
dimension $h_{5 5}$, $h_{0 5}$ and $h_{i 5}$ vanished, while their
transformation do not vanish. In order to bring any coordinate
system into a GN coordinate system, one has to choose $\xi^{0}$,
$\xi^{i}$ and $\xi^{5}$ appropriately to adjust the position of
the brane and to set $h_{55}$, $h_{0 5}$ and $h_{i 5}$ to zero.
The remaining gauge freedoms for the GN system make it possible to
redefine the coordinates inside the brane worldsheet. This
discussion is also valid for our anisotropic case now.

The spatial vector $\xi^{i}$ can be decomposed further into
\begin{equation}
\xi^{i} = \nabla^{i} \xi + \bar{\xi}^{i}.
\end{equation}
Here the vector $\bar{\xi^{i}}$ is transverse, i.e. $\nabla_{i}
\bar{\xi}^{i} = 0$ and the index is upper and lower by the
anisotropic spatial $3 \times 3$ metric. With this decomposition,
from the transformations of the bulk metric perturbation
(\ref{metricperturbationtransformation}), one can easily find the
transformations for the perturbation fields respectively.

\subsubsection{Scalar gauge transformation}

The scalar perturbations transform as
\begin{eqnarray}
\psi & \rightarrow & \tilde{\psi} = \psi - \dot{\xi}^{0} -
\frac{\dot{n}}{n} \xi^{0} - \frac{n^{\prime}}{n} \xi^{5}, \\
R & \rightarrow & \tilde{R} = R, \\
E & \rightarrow & \tilde{E} = E, \\
B & \rightarrow & \tilde{B} = B + n^{2} \xi^{0} - \xi.
\end{eqnarray}
\subsubsection{Vector gauge transformation}

The vector perturbations transform as
\begin{eqnarray}
F_{i} & \rightarrow & \tilde{F}_{i} = F_{i}, \\
S_{i} & \rightarrow & \tilde{S}_{i} = S_{i} + \dot{\bar{\xi}}_{i}.
\end{eqnarray}

\subsubsection{Tensor gauge transformation}

The tensor perturbation transforms as
\begin{equation}
s_{i j} \rightarrow \tilde{s}_{i j} = s_{i j} - \frac{2}{a_{i}
a_{j}} \nabla_{i} \nabla_{j} \xi - \frac{1}{a_{i} a_{j}} \left (
\nabla_{i} \bar{\xi}_{j} + \nabla_{j} \bar{\xi}_{i} \right ) - 2
\left ( \frac{\dot{a}_{i}}{a_{i}} \xi^{0} +
\frac{a^{\prime}_{i}}{a_{i}} \xi^{5} \right ) \delta_{i j}
\end{equation}

In the above transformation for the perturbation fields, we see
that they are very different from the isotropic case. In the
isotropic braneworld cosmological perturbation, the tensor
perturbation fields remain unchanged under the gauge
transformation. While here we see that the tensor perturbation
fields changed under gauge transformation, and the scalar and
vector perturbation fields in the $3 \time 3$ spatial perturbed
metric remain unchanged. This can be interpreted as the
anisotropic effect in the braneworld cosmological perturbation. In
the braneworld cosmological perturbation in an anisotropic
background bulk geometry, the anisotropic effect can be reflected
in the gauge transformations of the perturbation fields. Different
gauge choice will give different anisotropic effect. Since the
tensor modes of the cosmological perturbations are the
gravitational waves which can propagate independently, the
anisotropic effect must can be reflected in the evolution of the
cosmological gravitational waves.

In the following we will restrict ourselves only to consider the
system inside a subset of GN coordinate system, then $\xi^{5} = 0$
and $\xi^{0}$, $\xi$ and $\bar{\xi}^{i}$ are not depend on the
extra dimension.

\subsection{Perturbed Einstein equations in the bulk}

The evolution of the bulk metric perturbation is governed by the
perturbed Einstein equations in the bulk. Outside the embedded
brane, the Einstein equations read
\begin{equation}
R_{A B} = \kappa^{2} \left ( \check{T}_{A B} - \frac{1}{3}
\check{T} g_{A B} \right ),
\end{equation}
where $\check{T}_{A B}$ is the energy-momentum tensor of the bulk
matter and $\check{T} = \sum_{A B} g^{A B} \check{T}_{A B}$. Then
the form of the perturbed Einstein equations in the bulk are
\begin{equation}
\delta R_{A B} = \kappa^{2} \left ( \delta \check{T}_{A B} -
\frac{1}{3} \delta \check{T} g_{A B} - \frac{1}{3} \check{T} h_{A
B} \right ).
\end{equation}

In previous section we have reviewed the solutions for the
braneworld in an anisotropic bulk background. In this solutions,
the empty bulk with a cosmological constant is of particular
important. The perturbed Einstein equations for a vacuum bulk with
cosmological constant $\Lambda$ are simply
\begin{equation}
\delta R_{A B} = \frac{2}{3} \Lambda h_{A B}.
\end{equation}

With the definition of the metric perturbation $h_{A B}$, the
general form of the curvature perturbations can be expressed as
\begin{equation} \label{perturbedcurvature}
\delta R_{A B} = - \frac{1}{2} D_{A} D_{B} h - \frac{1}{2}
\sum_{C} D^{C} D_{C} h_{A B} + \frac{1}{2} \sum_{C} D^{C} D_{A}
h_{B C} + \frac{1}{2} \sum_{C} D^{C} D_{B} h_{A C},
\end{equation}
where $h$ denotes the trace of the metric perturbation, namely
\begin{equation}
h = \sum_{A B} g^{A B} h_{A B}.
\end{equation}
Now substituting the definition of the metric perturbation
(\ref{perturbed_metric}) into (\ref{perturbedcurvature}), after
some tedious but straightforward calculations, we can obtain the
explicit expressions of the curvature perturbation which can also
be separated into scalar, vector and tensor parts.

\subsubsection{Scalar components}

\begin{eqnarray}
\delta R^{S}_{0 0} & = & - 3 \ddot{R} - \triangle \ddot{E} +
\sum_{i} \frac{\dot{a}_{i}}{a_{i}} \dot{\psi} + \left ( 3
\frac{\dot{n}}{n} - 2 \sum_{i} \frac{\dot{a}_{i}}{a_{i}} \right )
\dot{R} + \frac{\dot{n}}{n} \triangle \dot{E} - 2 \sum_{i}
\frac{\dot{a}_{i}}{a_{i}} \dot{E}_{, i i} \nonumber\\
& & + \sum_{i} \frac{1}{a_{i}^{2}} \dot{B}_{, i i} -
\frac{\dot{n}}{n} \sum_{i} \frac{1}{a_{i}^{2}} B_{, i i} +
n^{2} \sum_{i} \frac{1}{a_{i}^{2}} \psi_{, i i} \nonumber\\
& & + n^{2} \left [ \psi^{\prime \prime} + \left ( 2
\frac{n^{\prime}}{n} + \sum_{i} \frac {a^{\prime}_{i}}{a_{i}}
\right ) \psi^{\prime} + 3 \frac{n^{\prime}}{n} R^{\prime} +
\frac{n^{\prime}}{n} \triangle E^{\prime} + \left ( 2
\frac{n^{\prime \prime}}{n} + 2 \frac{n^{\prime}}{n} \sum_{i}
\frac{a^{\prime}_{i}}{a_{i}} \right ) \psi \right ] \\
\nonumber\\
\delta R^{S}_{0 i} & = & - 2 \dot{R}_{, i} + \left ( 3
\frac{\dot{a}_{i}}{a_{i}} -  \sum_{j} \frac{\dot{a}_{j}}{a_{j}}
\right ) R_{, i} + \left ( \sum_{j} \frac{\dot{a}_{j}}{a_{j}} -
\frac{\dot{a}_{i}}{a_{i}} \right ) \psi_{, i} \nonumber\\
& & - \frac{1}{2} B^{\prime \prime}_{, i} + \frac{1}{2} \left (
\frac{n^{\prime}}{n} + 2 \frac{a^{\prime}_{i}}{a_{i}} - \sum_{j}
\frac{a^{\prime}_{j}}{a_{j}} \right ) B^{\prime}_{, i} \nonumber\\
& & + \left [ \frac{1}{n^{2}} \left ( \frac{\ddot{a}_{i}}{a_{i}} +
\frac{\dot{a}_{i}}{a_{i}} \left ( \sum_{j}
\frac{\dot{a}_{j}}{a_{j}} - \frac{\dot{a}_{i}}{a_{i}} \right ) -
\frac{\dot{n} \dot{a}_{i}}{n a_{i}} \right ) - 2 \frac{n^{\prime}
a^{\prime}_{i}}{n a_{i}} \right ] B_{, i} \nonumber\\
& & + \sum_{j} \left \{ \left ( \frac{a_{i}}{a_{j}} - 1 \right )
\dot{E}_{, jji} + \left [ \frac{\dot{a}_{j}}{a_{j}} \left (
\frac{a_{i}}{a_{j}}- 1 \right ) + \frac {\dot{a}_{i}}{a_{i}} -
\frac{\dot{a}_{i}}{a_{j}} \right ] E_{, jji} \right \} \\
\nonumber\\
\delta R^{S}_{i j} & = & \left \{ \frac{a_{i}^{2}}{n^{2}} \left [
\ddot{R} + \left ( 3 \frac{\dot{a}_{i}}{a_{i}} + \sum_{k}
\frac{\dot{a}_{k}}{a_{k}} - \frac{\dot{n}}{n} \right ) \dot{R} -
\frac{\dot{a}_{i}}{a_{i}} \dot{\psi} + \frac{\dot{a}_{i}}{a_{i}}
\triangle \dot{E} - \frac{\dot{a}_{i}}{a_{i}} \sum_{k}
\frac{1}{a_{k}^{2}} B_{, kk} \right . \right . \nonumber\\
& & \left . \left . + 2 \left ( \frac{\dot{n} \dot{a}_{i}}{n
a_{i}} - \frac{\dot{a}_{i}}{a_{i}} \left ( \sum_{k}
\frac{\dot{a}_{k}}{a_{k}} - \frac{\dot{a}_{i}}{a_{i}} \right ) -
\frac{\ddot{a}_{i}}{a_{i}} \right ) \psi \right ] - a_{i}^{2}
\left [ R^{\prime \prime} +
\frac{a^{\prime}_{i}}{a_{i}} \psi^{\prime} \right . \right .
\nonumber\\
& & \left . \left . + \left ( 3 \frac{a^{\prime}_{i}}{a_{i}} +
\sum_{k} \frac{a^{\prime}_{k}}{a_{k}} + \frac{n^{\prime}}{n}
\right ) R^{\prime} + \frac{a^{\prime}_{i}}{a_{i}} \triangle
E^{\prime} + \sum_{k} \frac{1}{a_{k}^{2}} R^{, kk} \right ] \right
\} \delta_{i j} + a_{i}^{2} {\cal M}_{i i} R \delta_{i j}
\nonumber\\
& & + {\partial_{i}
\partial_{j}} \left \{ a_{i} a_{j} \left [\frac{1}{n^{2}} \left (
\ddot{E} + \left ( \sum_{k} \frac{\dot{a}_{k}}{a_{k}} -
\frac{\dot{n}}{n} \right ) \dot{E} \right ) - \left ( E^{\prime
\prime} + \left ( \sum_{k} \frac{a^{\prime}_{k}}{a_{k}} +
{\frac{n^{\prime}}{n}} \right ) E^{\prime} \right ) \right ]
\right . \nonumber\\
& & \left . + a_{i} a_{j} {\cal M}_{i j} E + a_{i} a_{j} \left [
\sum_{k} \left ( \frac{1}{a_{i} a_{k}} + \frac{1}{a_{j} a_{k}} -
\frac{1}{a_{k}^{2}} \right ) E_{, kk} - \frac{1}{a_{i} a_{j}}
\Delta E \right ] - \psi - R \right . \nonumber\\
& & \left . + \frac{1}{n^{2}} \left [ - \dot{B} + \left (
\frac{\dot{n}}{n} + \frac{\dot{a}_{i}}{a_{i}} +
\frac{\dot{a}_{j}}{a_{j}} - \sum_{k}
\frac{\dot{a}_{k}}{a_{k}} \right ) B \right ] \right \} \\
\nonumber\\
\delta R^{S}_{i 5} & = & - \psi^{\prime}_{, i} - 2 R^{\prime}_{,
i} + \left ( \frac{a^{\prime}_{i}}{a_{i}} - \frac{n^{\prime}}{n}
\right ) \psi_{, i} + \left ( 3 \frac{a^{\prime}_{i}}{a_{i}} -
\sum_{j} \frac{a^{\prime}_{j}}{a_{j}} \right ) R_{, i}
\nonumber\\
& & + \frac{1}{n^{2}} \left \{ - \frac{1}{2} \dot{B}^{\prime}_{,
i} + \frac{a^{\prime}_{i}}{a_{i}} \dot{B}_{, i} + \frac{1}{2}
\left ( \frac{\dot{n}}{n} - \sum_{j} \frac{\dot{a}_{j}}{a_{j}}
\right ) B^{\prime}_{, i} \right . \nonumber\\
& & \left . + \left [ \frac{\dot{a}^{\prime}_{i}}{a_{i}} +
\frac{a^{\prime}_{i}}{a_{i}} \left ( \sum_{j}
\frac{\dot{a}_{j}}{a_{j}} - \frac{\dot{a}_{i}}{a_{i}} \right ) -
\frac{\dot{n} a^{\prime}_{i}}{n a_{i}} \right ] B_{, i} \right \}
\nonumber\\
& & + \sum_{j} \left \{ \left ( \frac{a_{i}}{a_{j}} - 1 \right )
E^{\prime}_{, ijj} + \left [ \frac{a^{\prime}_{i}}{a_{i}} +
\frac{a^{\prime}_{j}}{a_{j}} \left ( \frac{a_{i}}{a_{j}} - 1
\right ) - \frac{a^{\prime}_{i}}{a_{j}} \right ] E_{, ijj} \right
\} \\
\nonumber\\
\delta R^{S}_{5 5} & = & - \psi^{\prime \prime} - 3 R^{\prime
\prime} - \triangle E^{\prime \prime} - 2 \frac{n^{\prime}}{n}
\psi^{\prime} - 2 \sum_{i} \frac{a^{\prime}_{i}}{a_{i}} R^{\prime}
- 2 \sum_{i} \frac{a^{\prime}_{i}}{a_{i}} E^{\prime}_{, ii} \\
\nonumber\\
\delta R^{S}_{0 5} & = & - 3 \dot{R}^{\prime} - \triangle
\dot{E}^{\prime} + \sum_{i} \left ( \frac{n^{\prime}}{n} -
\frac{a^{\prime}_{i}}{a_{i}} \right ) \left ( \dot{R} + \dot{E}_{,
ii} \right ) + \sum_{i} \frac{\dot{a}_{i}}{a_{i}} \left (
\psi^{\prime} - R^{\prime} - E^{\prime}_{, ii} \right )
\nonumber\\
& & + \frac{1}{2} \sum_{i} \frac{1}{a_{i}^{2}} B^{\prime}_{, ii} -
\frac{n^{\prime}}{n} \sum_{i} \frac{1}{a_{i}^{2}} B_{, ii}
\end{eqnarray}

\subsubsection{Vector components}

\begin{eqnarray}
\delta R^{V}_{0 0} & = &  - 2 \sum_{i} \frac{\dot{a}_{i}}{a_{i}}
\dot{F}_{i, i} \\
\nonumber\\
\delta R^{V}_{0 i} & = & + \sum_{j} \left \{ \frac{a_{i}}{2 a_{j}}
\dot{F}_{i, jj} + \frac{a_{i}}{2 a_{j}} \dot{F}_{j, ji} +
\frac{1}{2} \left ( \frac{a_{i} \dot{a}_{j}}{a_{j}^{2}} -
\frac{\dot{a}_{i}}{a_{j}} \right ) F_{i, jj} \right. \nonumber\\
& & \left . + \left [ \frac{\dot{a}_{j}}{a_{j}} \left (
\frac{a_{i}}{2 a_{j}} - 1 \right ) + \frac{\dot{a}_{i}}{a_{i}}  -
\frac{\dot{a}_{i}}{2 a_{j}} \right ] F_{j, ji} \right \}
\nonumber\\
& & + \frac{1}{2} S^{\prime \prime}_{i} - \frac{1}{2} \left (
\frac{n^{\prime}}{n} + 2 \frac{a^{\prime}_{i}}{a_{i}} - \sum_{j}
\frac{a^{\prime}_{j}}{a_{j}} \right ) S^{\prime}_{i} \nonumber\\
& & - \left [ \frac{1}{n^{2}} \left ( \frac{\ddot{a}_{i}}{a_{i}} +
\frac{\dot{a}_{i}}{a_{i}} \left ( \sum_{j}
\frac{\dot{a}_{j}}{a_{j}} - \frac{\dot{a}_{i}}{a_{i}} \right ) -
\frac {\dot{n} \dot{a}_{i}}{n a_{i}} \right ) - 2 \frac{n^{\prime}
a^{\prime}_{i}}{n a_{i}} \right ] S_{i} + \frac{1}{2} \sum_{j}
\frac{1}{a_{j}^{2}} S_{i, jj} \\
\nonumber\\
\delta R^{V}_{i j} & = & \frac{a_{i} a_{j}}{2}  \sum_{k} \left [
\frac{1}{a_{j} a_{k}} \left ( F_{i, jkk} + F_{k, kij} \right ) +
\frac{1}{a_{i} a_{k}} \left ( F_{j, ikk} + F_{k, kji} \right ) -
\frac{1}{a_{k}^{2}} \left ( F_{i, jkk} + F_{j, ikk} \right )
\right ]
\nonumber\\
& &  + \frac{a_{i} a_{j}}{2} \left \{ {\cal M}_{i j} \left ( F_{i,
j} + F_{j, i} \right ) + \frac{1}{n^{2}} \left [ \ddot{F}_{i, j} +
\ddot{F}_{j, i} + \left ( \sum_{k} \frac{\dot{a}_{k}}{a_{k}} -
\frac{\dot{n}}{n} \right ) \left ( \dot{F}_{i, j} + \dot{F}_{j, i}
\right ) \right ] \right .
\nonumber\\
& & \left . \hspace{.5cm} - \left [ F^{\prime \prime}_{i, j} +
F^{\prime \prime}_{j, i} + \left ( \sum_{k}
\frac{a^{\prime}_{k}}{a_{k}} + \frac{n^{\prime}}{n} \right ) \left
( F^{\prime}_{i, j} + F^{\prime}_{j, i} \right )\right ] \right \}
\nonumber\\
& & + \frac{1}{2 n^{2}} \left [ \dot{S}_{i, j} + \dot{S}_{j, i} +
\left ( \sum_{k} \frac{\dot{a}_{k}}{a_{k}} - 2
\frac{\dot{a}_{j}}{a_{j}} - \frac{\dot{n}}{n} \right ) S_{i, j} +
\left ( \sum_{k} \frac{\dot{a}_{k}}{a_{k}} - 2
\frac{\dot{a}_{i}}{a_{i}} - \frac{\dot{n}}{n} \right ) S_{j, i}
\right ] \\
\nonumber\\
\delta R^{V}_{i 5} & = & + \sum_{j} \left \{ \frac{a_{i}}{2 a_{j}}
F^{\prime}_{i, jj} + \frac{a_{i}}{2 a_{j}} F^{\prime}_{j, ji} +
\frac{1}{2} \left ( \frac{a_{i} a^{\prime}_{j}}{a_{j}^{2}} -
\frac{a^{\prime}_{i}}{a_{j}} \right ) F_{i, jj} \right.
\nonumber\\
& & \left . + \left [ \frac{a^{\prime}_{i}}{a_{i}} +
\frac{a^{\prime}_{j}}{a_{j}} \left ( \frac{a_{i}}{2 a_{j}} - 1
\right ) - \frac{a^{\prime}_{i}}{2 a_{j}} \right ] F_{j, ji}
\right \}
\nonumber\\
& & - \frac{1}{n^{2}} \left \{ - \frac{1}{2} \dot{S}^{\prime}_{i}
+ \frac{a^{\prime}_{i}}{a_{i}} \dot{S}_{i} + \frac{1}{2} \left (
\frac{\dot{n}}{n} - \sum_{j} \frac{\dot{a}_{j}}{a_{j}} \right )
S^{\prime}_{i} \right . \nonumber\\
& & \left . + \left [ \frac{\dot{a}^{\prime}_{i}}{a_{i}} +
\frac{a^{\prime}_{i}}{a_{i}} \left ( \sum_{j}
\frac{\dot{a}_{j}}{a_{j}} - \frac{\dot{a}_{i}}{a_{i}} \right ) -
\frac{\dot{n} a^{\prime}_{i}}{n a_{i}} \right ] S_{i} \right \}
\\
\nonumber\\
\delta R^{V}_{5 5} & = &  - 2 \sum_{i}
\frac{a^{\prime}_{i}}{a_{i}} F^{\prime}_{i, i} \\
\nonumber\\
\delta R^{V}_{0 5} & = & - \sum_{i} \frac{a^{\prime}_{i}}{a_{i}}
\dot{F}_{i, i} - \sum_{i} \frac{\dot{a}_{i}}{a_{i}} F^{\prime}_{i,
i} - \frac{1}{2} \sum_{i} \frac{1}{a_{i}^{2}} S^{\prime}_{i, i} \\
\end{eqnarray}

\subsubsection{Tensor components}

\begin{eqnarray}
\delta R^{T}_{0 0} & = & - \sum_{i} \frac{\dot{a}_{i}}{a_{i}}
\dot{s}_{ii} \\
\nonumber\\
\delta R^{T}_{0 i} & = & + \frac{1}{2} \sum_{j}  \left [
\frac{a_{i}}{a_{j}} \dot{s}_{ij, j} + \left (
\frac{\dot{a}_{i}}{a_{i}} - \frac{\dot{a}_{j}}{a_{j}} \right )
s_{jj, i} + \left ( \frac{a_{i} \dot{a}_{j}}{a_{j}^{2}} -
\frac{\dot{a}_{i}}{a_{j}} \right )
s_{ij, j} \right ] \\
\nonumber\\
\delta R^{T}_{i j} & = & - \frac{1}{2} a_{i} a_{j} \sum_{k}
\frac{1}{a_{k}^{2}} s_{ij, kk} + \frac{1}{2} \frac{a_{i}
a_{j}}{n^{2}} \left [ \ddot{s}_{ij} + \left ( \sum_{k}
\frac{\dot{a}_{k}}{a_{k}} - \frac{\dot{n}}{n} \right )
\dot{s}_{ij} \right ] \nonumber\\
& & - \frac{a_{i} a_{j}}{2} \left [ s^{\prime \prime}_{ij} + \left
( \sum_{k} \frac{a^{\prime}_{k}}{a_{k}} + \frac{n^{\prime}}{n}
\right ) s^{\prime}_{ij} \right ] + \frac{a_{i} a_{j}}{2} {\cal
M}_{i j} s_{ij} \\
\nonumber\\
\delta R^{T}_{i 5} & = & + \frac{1}{2} \sum_{j} \left [
\frac{a_{i}}{a_{j}} s^{\prime}_{ij, j} + \left (
\frac{a^{\prime}_{i}}{a_{i}} - \frac{a^{\prime}_{j}}{a_{j}} \right
) s_{jj, i} + \left ( \frac{a_{i} a^{\prime}_{j}}{a_{j}^{2}} -
\frac{a^{\prime}_{i}}{a_{j}} \right ) s_{ij, j} \right ] \\
\nonumber\\
\delta R^{T}_{5 5} & = & - \sum_{i}
\frac{a^{\prime}_{i}}{a_{i}} s^{\prime}_{ii} \\
\nonumber\\
\delta R^{T}_{0 5} & = & - \frac{1}{2} \sum_{i}
\frac{a^{\prime}_{i}}{a_{i}} \dot{s}_{ii} - \frac{1}{2} \sum_{i}
\frac{\dot{a}_{i}}{a_{i}} s^{\prime}_{ii} \\
\end{eqnarray}
Here $\triangle f = \sum_{i} f_{, ii}$. and the quantity ${\cal
M}_{i j}$ was defined as
\begin{eqnarray}
{\cal M}_{i j} & = & \frac{1}{n^{2}} \left [
\frac{\ddot{a}_{i}}{a_{i}} + \frac{\ddot{a}_{j}}{a_{j}} -
\frac{\dot{a}_{i}^{2}}{a_{i}^{2}} -
\frac{\dot{a}_{j}^{2}}{a_{j}^{2}} + 4 \frac{\dot{a}_{i}
\dot{a}_{j}}{a_{i} a_{j}} + \left ( \sum_{k}
\frac{\dot{a}_{k}}{a_{k}} - \frac{\dot{a}_{i}}{a_{i}} -
\frac{\dot{a}_{j}}{a_{j}} - \frac{\dot{n}}{n} \right ) \left (
\frac{\dot{a}_{i}}{a_{i}} + \frac{\dot{a}_{j}}{a_{j}} \right )
\right ]
\nonumber\\
& & - \left [ \frac{a^{\prime \prime}_{i}}{a_{i}} +
\frac{a^{\prime \prime}_{j}}{a_{j}} -
\frac{{a^{\prime}_{i}}^{2}}{a_{i}^{2}} -
\frac{{a^{\prime}_{j}}^{2}}{a_{j}^{2}} + 4 \frac{a^{\prime}_{i}
a^{\prime}_{j}}{a_{i} a_{j}} + \left ( \sum_{k}
\frac{a^{\prime}_{k}}{a_{k}} - \frac{a^{\prime}_{i}}{a_{i}} -
\frac{a^{\prime}_{j}}{a_{j}} + \frac{n^{\prime}}{n} \right ) \left
( \frac{a^{\prime}_{i}}{a_{i}} + \frac{a^{\prime}_{j}}{a_{j}}
\right ) \right ]
\end{eqnarray}
If we take $a_{1} = a_{2} = a_{3} = a$, i.e. we go back to the
isotropic cosmological perturbation, then the expressions we
obtained above can reduce to all the results obtained by Langlois
in \cite{Langlois2}.

\section{Perturbation of the junction conditions}

Having obtained the formalism of the metric perturbation in the
bulk, this section we will discuss the perturbation of the matter
in the brane.

In subsection \ref{brane&juncitoncondition} we have discussed the
junction conditions which govern the dynamics of the brane. In the
anisotropic background bulk geometry, the matter of the brane can
be some kind of the isotropic perfect fluid matter. In the
following we will mainly discuss the matter perturbation of this
kind of brane matter.

The energy-momentum tensor of the matter in the brane is
\begin{equation}
T^{\mu}_{\nu} = \delta(y) S^{\mu}_{\nu}.
\end{equation}
Here we assume that the brane is located at the position of $y =
0$. For the perfect fluid kind of matter, we have
\begin{equation}
S^{\mu}_{\nu} = (\rho + P)u^{\mu} u_{\nu} + P g^{\mu}_{\nu}.
\end{equation}
From the junction condition (\ref{junction}) with $Z_{2}$
symmetry, we can easily get the jump of the extrinsic curvature
across the brane
\begin{equation}
[K_{\mu \nu}] = - \kappa^{2} \left ( S_{\mu \nu} - \frac{1}{3} S
g_{\mu \nu} \right ).
\end{equation}
Then the perturbed junction conditions are
\begin{equation}
[\delta K_{\mu \nu}] = \kappa^{2} \left ( - S_{\mu \nu} +
\frac{1}{3} g_{\mu \nu} \delta S + \frac{1}{3} S h_{\mu \nu}
\right ).
\end{equation}

In case of the GN coordinate system, the extrinsic curvature can
be expressed simply as
\begin{equation}
K_{\mu \nu} = \frac{1}{2} \partial_{y} g_{\mu \nu}.
\end{equation}
Using this simple form of the extrinsic curvature, the perturbed
junction conditions become
\begin{equation} \label{perturbed_junction}
h^{\prime}_{\mu \nu}|_{y = 0^{+}} = \kappa^{2} \left ( - S_{\mu
\nu} + \frac{1}{3} g_{\mu \nu} \delta S + \frac{1}{3} S h_{\mu
\nu} \right ).
\end{equation}
From the normalization condition satisfied by the 4-velocity, we
can get the perturbations of the unit 4-velocity
\begin{equation}
\delta u^{\mu} = \left \{ - n^{-1} \psi, a_{i}^{-1} v^{i} \right
\}.
\end{equation}
Then the components of the perturbation of the energy-momentum
tensor are
\begin{eqnarray}
\delta S_{0 0} & = & n^{2} \delta \rho + 2 n^{2} \psi \rho, \\
\delta S_{0 i} & = & - \left ( \rho + P \right ) n a_{i} v_{i} +
P B_{i} \\
\delta S_{i j} & = & a_{i} \delta_{i j} \delta P + P h_{i j} +
\pi_{i j}, \\
\end{eqnarray}
where the index of $v^{i}$ is still lower and upper by the
anisotropic spatial 3-metric: $v_{i} = \sum_{j}
\frac{1}{a_{i}^{2}} \delta_{i j} v^{j}$, $\pi_{i j}$ is the
anisotropic stress tensor.

As in the metric perturbation, it is also possible to decompose
above expressions further into scalar, vector and tesnsor
components by using the decomposition
\begin{equation}
v^{i} = \nabla^{i} v + \bar{v}^{i}
\end{equation}
with the divergence-free condition for $\bar{v}^{i}$: $\sum_{i}
\nabla_{i} \bar{v}^{i} = 0$ and
\begin{equation}
\pi_{i j} = \left ( \nabla_{i} \nabla_{j} - \frac{1}{3} \sum_{k}
\frac{1}{a_{k}^{2}} \nabla_{k} \nabla_{k} \delta_{i j} \right )
E_{\pi} + \nabla_{i} F_{\pi j} + \nabla_{j} F_{\pi i} + s_{\pi i
j}
\end{equation}
with $F_{\pi i}$ anisotropic transverse and $s_{\pi i j}$
anisotropic transverse traceless.

Substituting these expressions and the background quantities into
(\ref{perturbed_junction}), we can get the explicit form of the
perturbed junction conditions directly.

\section{Conclusions}

In this paper, we have developed a formalism of the cosmological
perturbation for the anisotropic braneworld. First we reviewed the
solutions of the Einstein equations for the braneworld model in an
anisotropic background bulk geometry. Since the anisotropic
background bulk geometry can support a still anisotropic brane but
with perfect kind of matter in it, in the sense of brane
cosmology, this means that the homogeneous and isotropic matter
distributions of our universe do not imply our universe (the brane
embedded in the bulk) is necessarily isotropic.

Then we turned to calculate the cosmological perturbation of this
kind of anisotropic braneworld model. The anisotropic braneworld
model is natural generalization of the usual homogeneous and
isotropic braneworld model, the formalism of the cosmological
perturbation obtained in present paper is also a direct
generalization of the isotropic braneworld cosmological
perturbation. When we go back to the isotropic case by taking all
the spatial scale factor to be same: $a_{1} = a_{2} = a_{3} = a$,
we can recovery all the cosmological perturbation results for the
isotropic case.

The difference of the cosmological perturbation between the
anisotropic and isotropic braneworld models is caused by the
anisotropic effect reflected in the braneworld cosmological
perturbation. It is well known that the cosmological perturbation
plays a key role in the study of the CMBR, in the context of
braneworld with anisotropic background, the anisotropic effect can
also be reflected in the anisotropic spetrum of the CMBR. This
will be considered in our future work.

\begin{acknowledgements}
This work was partly supported by the NSFC, Grant No. 10347139 and
partly supported through USTC ICTS by grants from the Chinese
Academy of Science and a grant from NSFC.
\end{acknowledgements}

\end{document}